# The exoplanets analogy to the Multiverse

Osame Kinouchi[1]

Physics Department – FFCLRP – USP

**Abstract**

The idea of a Mutiverse is controversial, although it is a natural possible solution to particle physics and cosmological fine-tuning problems (FTPs). Here I explore the analogy between the Multiverse proposal and the proposal that there exist an infinite number of stellar systems with planets in a flat Universe, the Multiplanetverse. Although the measure problem is present in this scenario, the idea of a Multiplanetverse has predictive power, even in the absence of direct evidence for exoplanets that appeared since the 90´s. We argue that the fine-tuning of Earth to life (and not only the fine-tuning of life to Earth) could predict with certainty the existence of exoplanets decades or even centuries before that direct evidence. Several other predictions can be made by studying only the Earth and the Sun, without any information about stars. The analogy also shows that theories that defend that the Earth is the unique existing planet and that, at the same time, is fine-tuned to life by pure chance (or pure physical necessity from a parameter free Theory of Everything) are misguided, and alike opinions about our Universe are similarly delusional.

**Introduction**

This paper intends to be a direct response to a recent book chapter from Klaas Landsman entitled "The Fine-Tuning Argument" (Landsman, 2016). In that text Landsman argues that FTPs does not support the idea of a Mutiverse and that we need to accept our Universe as given, not considering the possibility of alternative physical laws and universal constants. That is, for Landsman, universal numbers like the cosmological constant $\Lambda$, the Higgs mass $m_H$, the electromagnetic constant $\alpha$, the gravitational contant $G$ and the ratio between u and d quarks mass (and, by consequence, the neutron/proton mass ratio) are not parameters to be tuned to obtain a complex[2] universe but must be accepted as given.

---

[1] okinouchi@gmail.com
[2] By complex Universe we denote a Universe with stars and chemistry.

It is difficult to understand what exactly this "given" means. Landsman list as solutions to the FTP the following:

1. *Design*: There is a creator that fine tuned the Universe to be hospitable to life.
2. *Multiverse*: There is a collection of Universes with different nature constants and ours is one of the few that have the fine-tuned parameters.
3. *Blind chance*: The universal numbers have the good values by pure chance.
4. *Blind necessity*: The universal numbers are uniquely determined by a deeper physical theory that has no free parameters, a Theory of Everything (TOE).
5. *Misguided*: the fine-tuning problem should be resolved by some appropriate therapy. We must accept the Universe as given.

This last option is the preferred by Landsman. However, option 5 is hardly distinguishable from 3 or 4. To accept the universal constants (and physical laws and initial conditions) as given is equivalent to assume that 3 or 4 are valid.

We will not discuss option 3 because it is hardly scientific. It requires that the Goddess of chance is more powerful that the God of the Design. After all, in religious traditions, the creator spends some time and do some work to create the Universe. Chance, however, creates the right Universe without effort or time lapse. Chance is more powerful than the omnipotent God. This could be possible, but is not scientifically credible.

So, I presume that option 5 is based in option 4 (blind necessity). However, by modern standards, the faith that a future TOE shall be a parameter free theory is precisely this: an act of faith. From our present knowledge, there is no hint that such theory is possible, and the most successful research programs like superstrings and inflation point that changes in these parameters and laws are even unavoidable (Wilczek, 2013; Schellekens, 2013; Susskind, 2012, Bousso, 2011; Greene, 2011; Weinberg, 2009). The single argument that Landsman has to contest these claims is to call these research programs as speculative, but we must remember that their competitors are even more speculative, have less explicative power and, in the case of inflation, have less empirical support (Guth *et al.*, 2014).

Landsman also argues, or better, states, that "it is not the Universe that is fine-tuned for life, but life that has been fine-tuned to the Universe". This is a somewhat misleading affirmation, first because both phrases are not logically incompatible: the Universe could be fine-tuned to life and, at the same time, life could be adapted to some parts of the Universe (some planets but not others). Second, the adaptive power of life is highly limited: we do not expect to find life in Mercury or Venus, and even Mars has no visible biosphere.

It is a truism that complex life evolved and adapted to Earth, but first one needs an Earth with the correct conditions and parameters to support life (to be described later). Variations in these parameters would lead to no evolution, or evolution only of microbial life (Ward and Brownlee, 2000; Forgan and Rice, 2010). So, biological evolution needs an ambient that favors, or at least permits it. Biological adaptation power does not explain why life evolved only in the Earth instead of in the Moon, nor why microbial life carried by Earth meteorites to the Moon did not evolve in a full biosphere there. Evolution has inherent limitations.

The fine tuning problem does not vanish by appealing to adaptive evolution. Indeed, Landsman affirmation seems to be not an argument but more a catch phrase, since he does not fully develop or evaluate the idea in his paper (Landsman, 2016). This being established, lets turn to the full planetary FTP.

### The Multiplanetverse analogy

Lets start with our analogy, or better, our thought experiment, which we call the Multiplanetverse hypothesis. Suppose that the atmosphere of Earth was filled by a permanent global fog, so that only the Sun and the Moon were visible but stars were invisible even to the most potent telescopes. We also assume that, at the given time, scientists were not able to send rockets and satellites to space. In such scenario, astronomers only can study the Moon, the Sun and the free fall of bodies. Although this is a very restricted situation, it is credible that such astronomers and physicists could elaborate a gravitation theory similar to that of Newton.

However, they start to wonder that the distance $D$ from the Sun to Earth and the orbit eccentricity $\varepsilon$ put it well inside the habitable zone where surface temperature allows liquid water and water-based life. The Sun mass $M_S$ also lies within a critical range (large enough to push the habitable zone outside the planet tidal locking radius, and small enough to provide sufficient energy while avoiding UV exposure). The Sun mass also enables constant energy production for billions of years needed by biological evolution. Also, they notice that the Earth radius $R$ and the Earth mass $M$ have values inside the parameter region that allows a suitable atmosphere.

The mass of planetary water $M_{H2O}$ is also sufficient to produce global oceans but with the right amount so that free lands are possible. This varied landscape seems also to favor the emergence of complex life (continents above the sea level indeed are a prerequisite to the appearance of technological animals). They call these observations "the fine-tuning problem" (FTP) because their gravitation theory (and other physical theories) cannot predict or explain these numbers.

Adepts of a parameter free TOE would argue that a more fundamental and deeper physical theory would furnish these numbers, so that the fine-tuning of $D$, $\varepsilon$, $M_S$, $R$, $M$, $M_{H2O}$, and other numbers, would be fully explained and determined by *ab initio* calculations. They propose that the FTP should be solved by therapy (Landsman, 2016) because it is an imaginary problem. However, no such TOE is proposed, there is no promising research program that aims to found such TOE, there is no hint that such TOE exists. Believing in such TOE is a pure act of faith.

So, what is the natural solution of the Earth FTP? Although stars and stellar planetary systems are out of reach to our hypothetical astronomers, the fact that they recognize that the set $P = \{D, \varepsilon, R, M_S, M, M_{H2O}\}$ cannot be derived from a TOE has a single outcome: they must be environmental and mostly random[3]. So, they make a bold and highly speculative hypothesis: that there is an infinite set of planets[4], each with different parameters $P$, so that Earth is one of the rare planets with the correct set that favors life. They call this bold conjecture as the Multiplanetverse. The fact that we inhabit just a planet like Earth, with a lot of fine-tuning, is not a surprise: otherwise we would not be here to make questions about fine-tuning. The Weak Anthropic Principle is fully applicable here (Weinstein, 2006; Schellekens, 2013).

However, the Multiplanetverse is open to criticism. First, it proposes almost any kind of planets, so it is not predictive: a theory that predicts everything predicts nothing. Also, we cannot derive a probability measure for $P$ since there is an infinite set of planets (Garriga and Vilenkin, 2013). Furthermore, to appeal to innumerous non observables planets to explain the single case of the Earth violates Occan Razor and abandons the Physics tradition of explaining things from first principles.

---

[3] Another hypothesis involves Design: a higher intelligence planned and created a unique planet, the Earth, with the right parameters to favor life. Landsman (2016) criticize the relation between FTP and Design and I find his arguments correct.

[4] Here we use and infinite set of planets by considering a flat and infinite Universe, with the same physical laws that lead to an infinite number of stars (Greene, 2011).

The fact that we cannot define a probability measure over *P* does not affect the fact that exoplanets exist. Nor the fact that the hypothesis is not predictive, in the sense that any fine tuning present in the Earth is also present in some fraction of planets of the Multiplanetverse. And the Occan Razor cuts in the wrong way the theory. So, these problems, also present in the Multiverse hypothesis, does not precludes that it be correct, like the Multiplanetverse hypothesis.

**Predictions from the Multipanetverse hypothesis**

By only studying a single exemplar, our planet, what our imaginary astronomers and physicists adept of the Multipanetverse hypothesis could infer about possible other planets? It is interesting that the same predictions could be made by historic astronomers even before the XVIII century (I do not know if such speculations were made at that epoch, and would be grateful for any reference given by readers).

I give bellow a non exhaustive list of predictions of the Multiplanetverse hypothesis:

1. There are innumerous stellar systems with a varied set of planets with different parameters *P*. This is needed to solve the planetary FTP and is a correct prediction (Laughlin and Lissauer, 2015).
2. Probably there are other rocky planets in the solar system. This prediction is motivated by the fact that it is more probable that one in, say, four planets justly fall in the habitable zone than that just a unique rocky planet falls inside that zone. This is a correct prediction.
3. In our hypothetical world, theoretical astrophysics (even without direct evidence of stars) can predict that red dwarfs are more abundant than Sun-like stars and have larger lifetime. From a typicality argument, this would lead to the conclusion that our star should be a red dwarf, other things being equal. But the Sun is not a red dwarf, so things are not equal: the prediction is that red dwarfs have a smaller number of habitable planets than Sun-like stars. Indeed this seems to be the case (Cohen *et al*., 2014).
4. Comte de Buffon theory of planetary formation, and similar theories that postulate that the solar system has been formed by a rare collision between the Sun and another star, is not compatible with most stars having planets, as proposed by the Multiplanetverse hypothesis. Indeed, such theories are wrong.

5. Since there are more moons (around giant planets) than planets, and supposing that some of these exomoons are habitable, our presence in a planet instead of a habitable moon needs explanation if we use an argument of typicality. The conclusion is that there are some factors that diminish the possibility of emergence of complex life in exomoons. These factors exist and are being investigated (Heller, 2012).

By combining the principle of mediocrity with the Rare Earth hypothesis we can derive other predictions. The idea is that what is rare in Earth is the joint probability of the set *P*, but not the individual factors by themselves, that are mediocre. So, our hypothetical astronomers, by examining only the Sun and the Earth, could predict by the principle of mediocre rarity[5] that:

6. There exist a lot of Sun-like stars.
7. There are a lot of planets whose *D* and *ε* falls inside the habitable zone.
8. There are a lot of planets with masses and radius similar to Earth.
9. Water is a common element in planetary surfaces.

So, without any information about stars and exoplanets, our astronomers can make these correct conjectures. With more exercise of imagination, other predictions could be derived from the Mutiplanetverse scenario. The important point here is not these specific predictions, but the fact that, like in the Multiverse scenario, we can make predictions by examining a single exemplar (our planet or our Universe), in a situation that does not need direct evidence of other planets or Universes. Importantly, such predictions can be made without solving the measure problem for planets or universes. Curiously, the principle of rare mediocrity seems to be valid for our Universe under the hypothesis of the Multiverse, that is, our Universe seems to lie in a dangerous region at the border of the parameter

---

[5] The principle of mediocre rarity is also valid for the Multiverse. There is a misunderstanding about fine tuning claims: that a parameter or factor is relevant to the fine tuning problem only if it is tightly constrained by anthropic considerations. However, gross tuning is also relevant to global fine tuning. For example, suppose that only ten of the nineteen parameters of the Standard Model must be tuned to give a Universe with chemistry and that the anthropic range for each occupy *1/10* of the allowed parameter range (a very gross tuning by any standard). Thus, the global fine tuning is *$10^{-10}$*. Simultaneous variation of parameters change a little this scenario, but does not alleviate the global fine tuning: suppose two covarying parameters $y \in [0, y_{max}]$ and $x \in [0, x_{max}]$ with the anthropic range having, say, a diagonal form *y = x ± Δy*, $y = x \pm \Delta y$. The allowed relative parameter volume $A_{xy}/(x_{max} \cdot y_{max})$ is of the same order of $\Delta y / y_{max}$ because the anthropic area is $A_{xy} \approx \Delta y \cdot x_{max}$.

space sector that allows complex life (Schellekens, 2013; Degrassi *et al.*, 2012).

**Conclusion**

As noticed by other authors, the difference between the Multiverse and the Multiplanetverse is not of kind but only of grade (Weinberg, 2009). Both postulate an infinite set of entities to explain the fine-tuning problem. In both cases, criticism by using Occan Razor is misguided, and the measure problem, although present, does not affect the concreteness or plausibility of such scenarios.

We also must remember that what makes a research program scientific is not only empirical support. In absence of that (a common situation in theoretical and frontier physics), we must evaluate ideas by their fertility, that is, their capacity of stimulate new research, new methods, new calculations, new ideas. In this sense of fertility, the Multiverse paradigm is very successful, as can be illustrated by Figure 1.

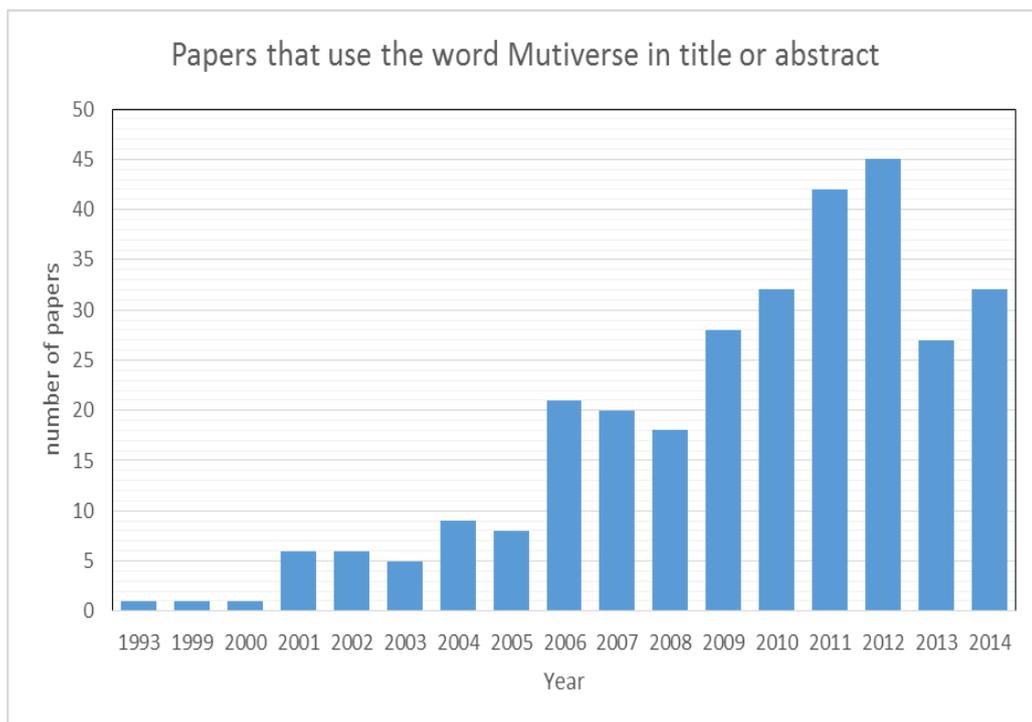

**Figure 1**: Number of papers from ArXiv that uses the word Multiverse in the title or abstract. Here are not counted several papers that use or discuss the Multiverse hypothesis only in the body text as, for example, Degrassi *et al.* (2012).

The Kepler dream of obtaining planetary orbit radius from first principles from a TOE is simply that: an unrealizable and unnecessary dream. People that think in the same way about the Universe fine-tuning problem need to eliminate such dreams and delusions. Since the crescent consensus between first rank scientists, even Nobels like Weinberg and Wilkzek, is that the Multiverse is a worth pursuing idea[6], and since the sceptics are a decaying minority, it seems that dreams are dissipating (Weinberg, 1994), and a scientific revolution is becoming.

*A scientific truth does not triumph by convincing its opponents and making them see the light, but rather because its opponents eventually die and a new generation grows up that is familiar with it.* Max Planck.

*Science advances one funeral at a time*. Paraphrased variant.

**Acknowledgments:** I thank very much Nestor Caticha for critical reading of the manuscript and giving valuable commentaries. Financial support from CNAIPS-USP and NEUROMAT-FAPESP is acknowledged.

---

[6] This is not an authority argument, but only a sociological observation that signalizes that we are experiencing a paradigm shift.